\documentclass[twocolumn,showpacs,floatfix]{revtex4}
\usepackage{graphicx}

\begin{document}

\title{Collective modes and ballistic expansion of a Fermi gas
in the BCS-BEC crossover}
\author{Hui Hu$^1$, A. Minguzzi$^1$, Xia-Ji Liu$^2$, and M. P. Tosi$^1$ }
\affiliation{ $^1$\ NEST-INFM and Classe di Scienze, Scuola
Normale Superiore, I-56126 Pisa, Italy \\
$^2$\ Department of
Physics, Tsinghua University, Beijing 100084, China }
\date{\today}

\begin{abstract}
We evaluate the frequencies of collective modes and the
anisotropic expansion rate of a harmonically trapped Fermi
superfluid at varying coupling strengths across a Feshbach
resonance driving a BCS-BEC crossover. The equations of motion for
the superfluid are obtained from a microscopic mean-field
expression for the compressibility and are solved within a scaling
ansatz. Our results confirm non-monotonic behavior in the
crossover region and are in quantitative agreement with current
measurements of the transverse breathing mode by Kinast {\it et
al.} [Phys. Rev. Lett. {\bf 92}, 150402 (2004)] and of the axial
breathing mode by Bartenstein {\it et al.} [Phys. Rev. Lett. {\bf
92}, 203201 (2004)].
\end{abstract}

 \maketitle

Current experiments on ultracold Fermi gases are rapidly advancing
towards the realization of superfluid states, and Bose-Einstein
condensation of dimers has already been achieved
\cite{jin04,grimm04,jin03,grimm03,ketterle,salomon}. A key tool
for the manipulation of atomic gases is the use of a Feshbach
resonance to vary the magnitude and sign of the coupling strength.
Across the resonance the $s$-wave scattering length $a$ goes from
large positive to large negative values, thus allowing exploration
of the crossover from the Bardeen-Cooper-Schrieffer (BCS) state to
the Bose-Einstein condensate (BEC) of bound-fermion pairs. As
Fermi gases have been demonstrated to be stable also near the
resonance \cite{thomas03}, they offer a new opportunity to
investigate highly correlated many-body systems.

As in the case of bosonic clouds, the frequencies of collective
modes of Fermi gases can be measured to high accuracy and can
yield information on the state of the system and on its
collisional properties, thus contributing to its characterization
in a strong-coupling regime. Another dynamical observable which is
readily accessed in the experiments is the aspect ratio of an
expanding cloud released from an anisotropic trap, which depends
on its quantum state and on its degree of collisionality. While a
normal Fermi gas in the collisionless regime expands spherically,
an anisotropic expansion is predicted for a normal gas in the
collisional regime as well as for a superfluid gas.

The purpose of this Letter is to present a theory of the
collective modes and of the expansion of a trapped superfluid
Fermi gas at zero temperature as the coupling strength is varied
across a Feshbach resonance. The frequencies of the collective
modes are well known in both the BCS and the BEC limit
\cite{stringari96,baranov,amoruso}. From non-mean-field
perturbative estimates it has been conjectured that the frequency
of the transverse breathing mode in a highly elongated trap should
exhibit a non-trivial dependence on the scattering length
\cite{stringari}. Two further studies have been based on
semi-empirical forms of the equation of state
\cite{heisenberg,kim}. Here we use a microscopic mean-field
description of the BCS-BEC crossover \cite{leggett,NSR}, which at
zero temperature is believed to capture the essential physics in
all regimes \cite{randeria}. We calculate the equation of state
and the density profiles of the gas under axially symmetric
confinement with the help of a local density approximation (LDA),
and use them to determine the collective mode frequencies and the
expansion rate by means of a simple scaling assumption. Our
results do not use an interpolation scheme nor involve adjustable
parameters, and show already at mean-field level non-monotonic
behaviors across a Feshbach resonance.

The frequencies of radial and axial breathing modes have very
recently been measured in elongated clouds of $^6$Li atoms at
various values of the coupling strength in the strong-coupling
intermediate regime \cite{kinast,grimm}. Quite remarkably, our
mean-field approach yields predictions for the frequencies of the
breathing modes that are in good quantitative agreement with these
experiments.

\noindent {\it Equation of state and equilibrium density profile}.
--- Our starting point is to determine the chemical potential $\mu
\left( n\right) $ as a function of density $n$ for a homogeneous
Fermi gas through the BCS-BEC crossover at zero temperature. We
use a mean-field theory proposed in the pioneering work of Leggett
\cite{leggett} and of Nozi\`{e}res and Schmitt-Rink \cite{NSR}.
This describes well both the BCS weak-coupling regime and the BEC
strong-coupling limit, and is believed to be a reliable
approximation in between \cite{leggett,NSR,randeria}. It has also
been applied  to a trapped gas within the LDA \cite{perali,lv}.

The gas consists of two equally populated spin components that
interact by a contact pseudopotential parametrized by the $s$-wave
scattering length $a$. The mean-field theory for the crossover
extends the usual BCS gap equation
\begin{equation}
1=\frac{4\pi \hbar ^2a}m\int \frac{d^3 k}{\left( 2\pi \right)
^3}\left[ \frac 1{2\epsilon _{{\bf k}}}-\frac 1{2E_{{\bf
k}}}\right]  \label{gap}
\end{equation}
and number equation
\begin{equation}
n =\int \frac{d^3 k}{\left( 2\pi \right) ^3}\left( 1-
\frac{\epsilon _{{\bf k}}-\mu }{E_{{\bf k}}}\right) \label{num}
\end{equation}
to all values of the pairing interaction. Here $ \epsilon _{{\bf
k}}=\hbar ^2 k^2/2m$, $E_{{\bf k}}=\sqrt{(\epsilon_{\bf k}-\mu
)^2+\Delta ^2}$, $\Delta $ is the gap parameter, and $n$ is the
total density for fermions in either spin state ($n_{\uparrow
}=n_{\downarrow }=n/2$). These equations are to be solved for $\mu
$ and $\Delta $ with a given choice of the dimensionless coupling
parameter $k_Fa\equiv \left( 3\pi ^2n \right) ^{1/3}a$. In the
weak-coupling limit ($k_Fa\rightarrow 0^{-}$) they give back the
standard BCS result, and when $ k_Fa\rightarrow 0^{+}$ they
correctly reproduce the binding energy of fermion pairs at leading
order \cite{randeria}. In the unitarity limit ($a=\pm \infty$) the
model gives $\mu(n)\propto n^{2/3}$, scaling as predicted by the
universality hypothesis \cite{ho2004}.

We then use the equation of state $\mu(n)$ to determine the
density profile $n_0({\mathbf r})$ under the confining potential
$V_{ext}({\mathbf r})=m \left [ \omega_\perp^2(x^2+y^2)+m
\omega_z^2 z^2 \right ]/2$ through the implicit LDA equation $\mu(
n_0({\mathbf r}))+V_{ext}({\mathbf r})=\mu_g$. Here $\mu_g$ is the
chemical potential of the trapped gas from the normalization
condition $N=\int d^3r\,n_0 \left( {\bf r}\right) $, $N$ being the
total number of fermionic atoms. A useful dimensionless coupling
parameter for a trapped gas is $\kappa =
(N^{1/6}a/a_{ho})^{-1}\approx 1.695(k^0_Fa)^{-1}$, where $k^0_F$
is the Fermi wave number at the trap center \cite{stringari}.

\begin{figure}[tbp]
\centerline{\includegraphics[width=6.0cm,angle=-90,clip=]{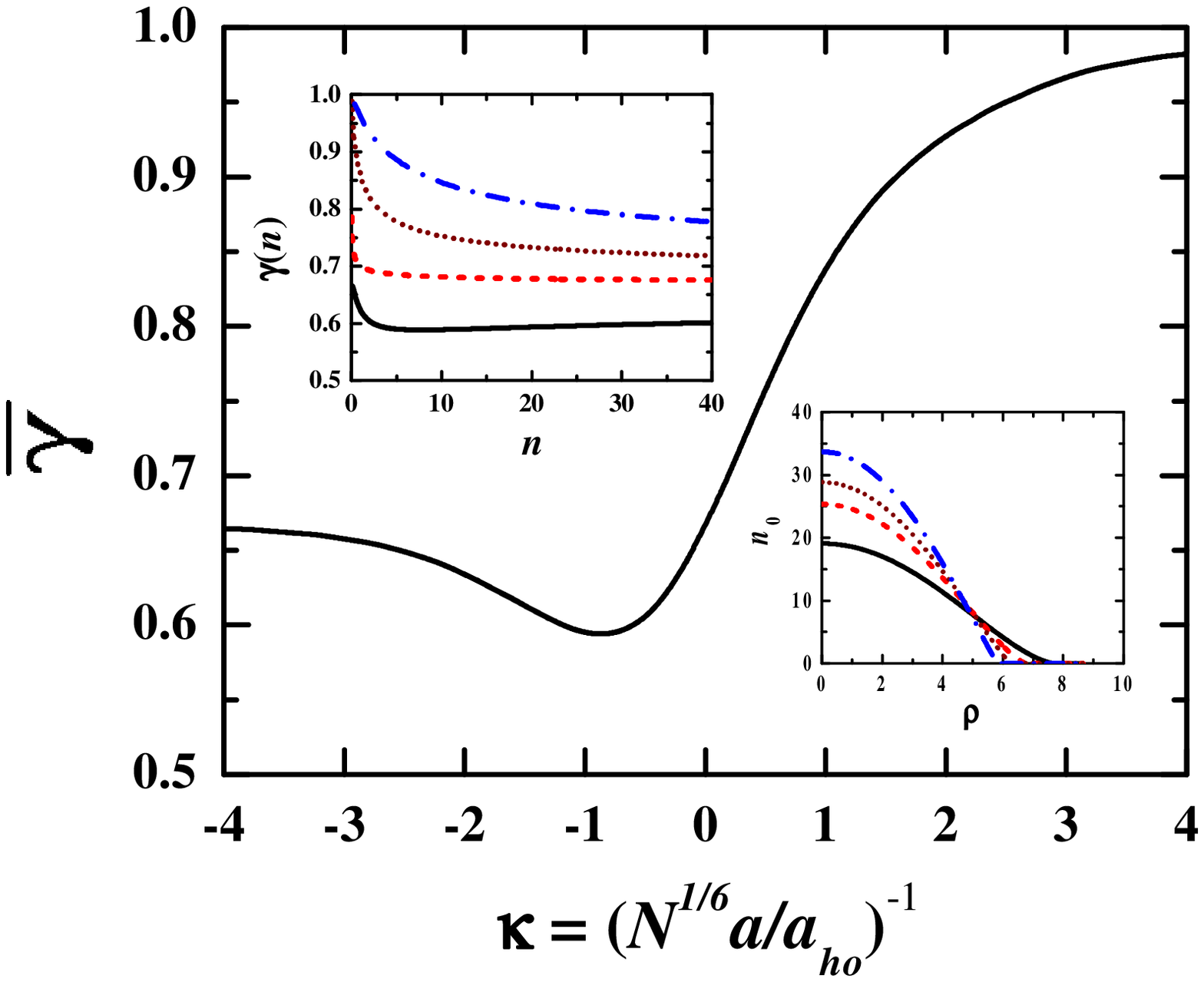}}
\caption{(color online) The exponent $\bar {\gamma}$ as a function
of the dimensionless coupling parameter
$\kappa=(N^{1/6}a/a_{ho})^{-1}$ for a trapped Fermi gas with $
N=2\times 10^5$ and $\lambda =\omega _z/\omega _{\bot }=0.05$. The
quantity $\gamma \left ( n \right )$ and the transverse density
profile $n_0 \equiv n_0 \left (\rho, 0\right )$ with $\rho =
\sqrt{x^2+y^2}$ are plotted in the insets for $ \kappa=-1.0$,
$+0.1$, $+0.5$ and $+1.0$ (from bottom to top). The coordinate $
\rho $ is in units of $a_{ho}=\lambda ^{-1/6} \sqrt{\hbar /m\omega
_{\bot }}$ and $n_0$ is in units of $a_{ho}^{-3}$.} \label{fig1}
\end{figure}

Our results are illustrated in Fig. 1. The left inset reports the
function $\gamma(n)=1+n\left[ \left( d^2\mu /dn^2\right) /\left(
d\mu /dn\right) \right]$ in the homogeneous gas for four choices
of $\kappa$, while the right inset shows the LDA density profiles
in the transverse direction. In the special case of a power-law
dependence of the chemical potential on density $\mu \left(
n\right) \propto n^\gamma $, $\gamma (n)\equiv \gamma $. In the
general situation $\gamma (n)$ acts as an effective exponent. The
main body of Fig. 1 shows the quantity $\bar {\gamma} =\left( \int
d^3{\bf r}n_0r_\alpha ^2\gamma\left( n_0\right) \right) /\left(
\int d^3{\bf r}n_0r_\alpha ^2\right)$ with $\alpha =x,$ $y,$ or
$z$, to be used below as an averaged effective exponent for
inhomogeneous gas. In fact, in the present approximation this
exponent is independent of the index $\alpha$ and it is seen from
Fig. 1 that it decreases in going from the BEC to the BCS limit
--- {\it i.e.}, the trapped gas is relatively less compressible in
the BEC limit.

\noindent {\it Scaling ansatz}. --- We describe the dynamics of
density fluctuations in a superfluid Fermi gas by the equation of
continuity
\begin{equation}
\partial_t n+{\bf \nabla }\cdot \left( n{\bf v}\right) =0,
\end{equation} and the Euler
equation
\begin{equation}
m\partial_t {\bf v} +{\bf \nabla} \left[ \mu \left( n\right) +
V_{ext} + m{\bf v}^2/2\right] =0. \label{hydro2}
\end{equation}
Here $n\left( {\bf r},t\right) $ and ${\bf v}\left( {\bf r}
,t\right) $ are the time-dependent density profile and velocity
field of the trapped gas, $\mu \left( n\right) $ is the local
chemical potential expressed as a function of $ n\left( {\bf
r},t\right) $  in an adiabatic LDA, and $V_{ext}\left( {\bf
r},t\right) =m \left [ \omega _{\bot }^2\left( t\right) \left(
x^2+y^2\right) +\omega _z^2\left( t\right) z^2 \right ] / 2$ is the
axially symmetric trapping potential. The above equations apply as
well to a normal Fermi gas in the collisional regime. The
difference from a superfluid should emerge from some further
knowledge of the collisional behavior of the cloud or from its
response under rotation.

We adopt the following scaling form of the time-dependent density
profile,
\begin{equation}
n\left( {\bf r},t\right) =\left [ \prod_\alpha b_\alpha \left(
t\right) \right ]^{-1} n_0\left( x/b_x \left( t\right), y/b_y
\left( t\right), z/b_z \left( t\right) \right) \label{dsty}
\end{equation}
where $n_0\left( {\bf r}\right) $ is again the equilibrium density
profile and  the dependence of the profile $n\left( {\bf
r},t\right)$ on time is entirely contained in the scaling
parameters $b_\alpha \left( t\right) $ \cite{kagan96}. This ansatz
correctly reflects the expansion and compression along each axis
and therefore is appropriate for the description of the breathing
modes and of the ballistic expansion of the cloud. The
corresponding form of the velocity field is fixed by the equation
of continuity as $v_\alpha \left( {\bf r},t\right)
=\dot{b}_\alpha(t) r_\alpha /b_\alpha (t) $.

The scaling ansatz in Eq. (\ref{dsty}) is an exact solution of the
equations of motion if the equation of state is a power law $\mu
\left( n\right) \propto n^\gamma $. In this case the scaling
parameters obey the coupled differential equations
\begin{equation}
\ddot{b}_\alpha +\omega _\alpha ^2b_\alpha -\left ( \omega_\alpha
^2 / b_\alpha \right ) \left ( \prod_\beta b_\beta \right
)^{-\gamma} =0. \label{spower}
\end{equation}
However, with a more general form of the equation of state the
scaling solution is not satisfied at every position ${\mathbf r}$.
A useful approximation is to take a spatial average with the
weight function $r^2_\alpha n_0$ \cite{note_scaling}. After some
straightforward algebra we obtain
\begin{eqnarray}
&&\ddot{b}_\alpha +\omega _\alpha ^2b_\alpha -\frac{\omega
_\alpha ^2 / b_\alpha }{\left( \prod_\beta b_\beta \right) }\times   \nonumber \\
&&\left. \frac{\int d^3{\bf r}n_0r_\alpha ^2\left[ \left( d\mu
/dn\right) _{n=n_0 (\prod_\beta b_\beta)^{-1} }/\left( d\mu
/dn\right) _{n=n_0}\right] }{\int d^3{\bf r}n_0r_\alpha
^2}=0\right., \label{sfinal}
\end{eqnarray}
which reduces to Eq. (\ref{spower}) if $\mu \left( n\right)
\propto n^\gamma $.

\noindent {\it Breathing modes} --- We evaluate the transverse and
axial breathing modes by taking periodic variations of the
trapping frequencies
$\omega_{\perp,z}(t)=\omega_{\perp,z}(1+\epsilon \cos(\omega t))$
with $\epsilon \ll 1$. Linearization of  Eq. (\ref{sfinal}) yields
the linearized form of Eq. (\ref{spower}) with $\gamma $ replaced
by $\bar{\gamma}$. One immediately finds three mode frequencies,
one of which lies at $\omega = \sqrt{2} \omega _{\bot }$
independently of the coupling strength and belongs to a surface
mode with projected angular momentum $m=\pm 2$. The other two mode
frequencies are given by
\begin{eqnarray}
\label{modefreqs} \omega^2_{\pm} / \omega _{\bot }^2 &=& 1+ \bar{
\gamma} +(2+\bar {\gamma})\lambda ^2 / 2
\nonumber \\
&&\pm \sqrt{\left[ 1+\bar {\gamma} +(2+\bar {\gamma})\lambda
^2/2\right] ^2-2\left( 2+3\bar {\gamma} \right) \lambda ^2},
\end{eqnarray}
where $\lambda =\omega _z/\omega_\perp$ and the  $\pm $ signs
refer to the transverse and axial mode, respectively.

\begin{figure}[tbp]
\centerline{\includegraphics[width=8.0cm,angle=-90,clip]{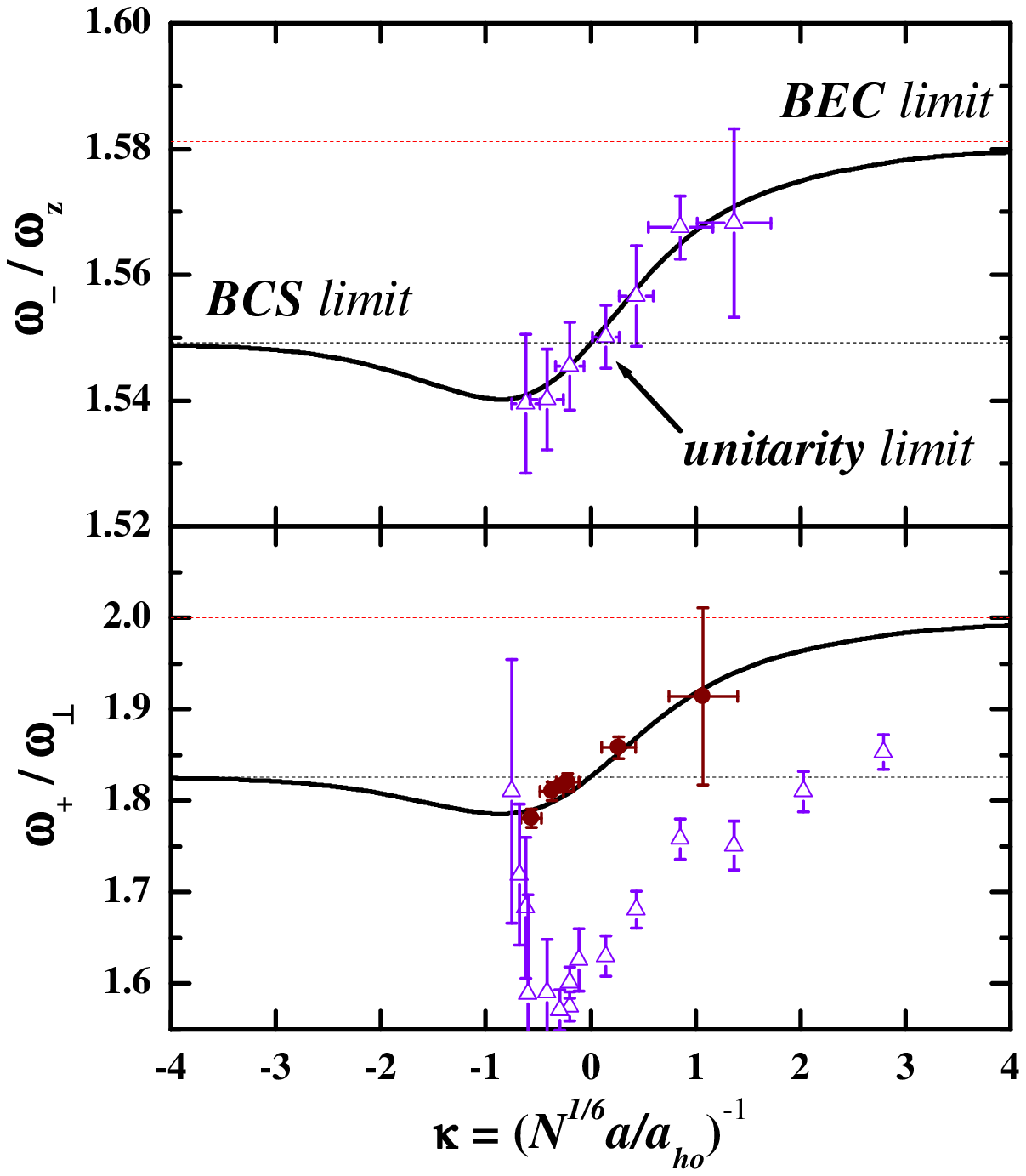}}
\caption{(color online) The frequency of the axial and transverse
breathing modes in an elongated trapped Fermi gas as functions of
$\kappa=(N^{1/6}a/a_{ho})^{-1}$. The solid circles and empty
triangles with error bars are the experimental results given by
Kinast {\it et al.} \cite{kinast} and by Bartenstein {\it et al.}
\cite{grimm}, respectively. Here we plot only the measured
frequencies with low dacay rates in the resonance region, for
instance, for the axial mode we use the data of the inset of
figure 1 in Ref. \cite{grimm}. The resonance position $B_0$ has
been set at 837 G. The error bar in the experimental data in the
horizontal ($\kappa$) axis is due to the uncertainty of $B_0$ and
we have taken $\Delta B_0 = \pm$ 15G.} \label{fig2}
\end{figure}

In the highly elongated traps of current experimental interest (
$\lambda \ll 1$) the frequencies of the breathing modes reduce to
$\omega _{+}/\omega _{\bot }=\sqrt{2+2\bar {\gamma}}$ and $\omega
_{-}/\omega _z=\sqrt{\left( 2+3\bar {\gamma} \right) /\left(
1+\bar {\gamma} \right) }$, yielding $\omega _{+}=$
$\sqrt{10/3}\omega _{\bot }\approx 1.83\omega _{\bot }$ in a
dilute Fermi gas ($\bar {\gamma} =2/3$) and $\omega _{+}=2\omega
_{\bot }$ in a dilute BEC ($\bar {\gamma} =1$). The mode
frequencies in the crossover region are presented in Fig. 2 as
solid lines displaying a non-trivial dependence on the coupling
parameter. In particular, a dip appears near the unitarity limit
on the BCS side, as a result of the pairing that enhances the
compressibility of the gas. On the BEC side the frequencies
monotonically increase towards the BEC value, this behavior being
related in our approximate scheme to the increase of the effective
exponent $\bar {\gamma}$ with the coupling constant $\kappa$ for
$\kappa>0$. In Fig. 2 we have also compared our predictions for
the axial mode with experimental data by Bartenstein {\it et al.}
\cite{grimm} and those for the transverse mode with measurements
by Kinast {\it et al.} \cite{kinast} and by Bartenstein {\it et
al.} \cite{grimm}. Two sets of the experimental data, {\it i.e.},
the axial mode in Ref. \cite{grimm} and the transverse mode in
Ref. \cite{kinast}, are in quantitative agreement with our
mean-field results.

\begin{figure}[tbp]
\centerline{\includegraphics[width=5.5cm,angle=-90,clip]{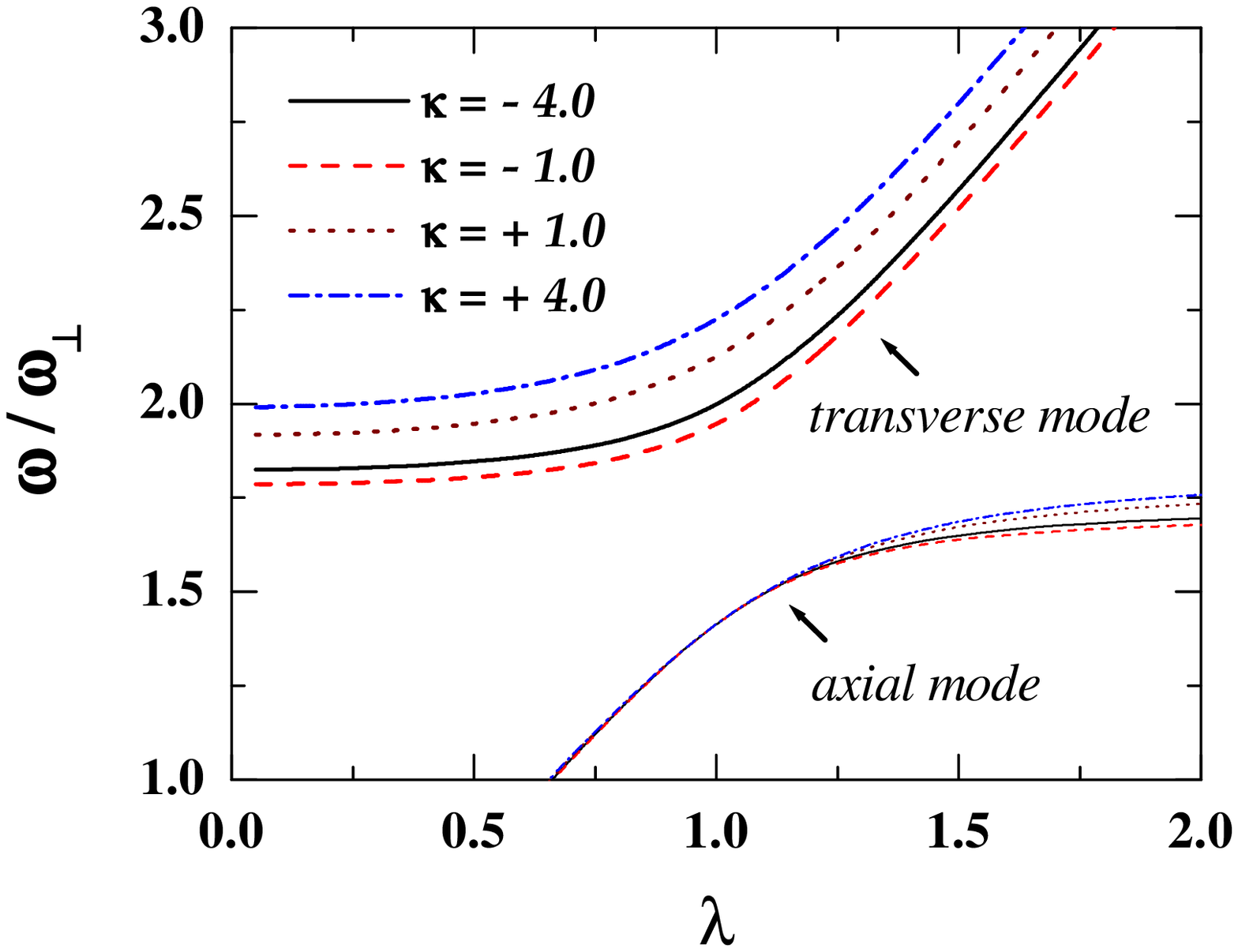}}
\caption{(color online) The frequency of the transverse and axial
breathing modes as functions of the anisotropy parameter $\lambda$
for a trapped Fermi gas with $N=2\times 10^5$, at the indicated
values of the coupling constant $\kappa$.} \label{fig3}
\end{figure}

However, the remaining set of data for the transverse mode
measured by Bartenstein {\it et al.} \cite{grimm} contradicts our
prediction. The measured frequencies have an abrupt change around
the resonance, accompanied by large damping rates. This is not
expected by the theory. As noted by Grimm, this might be due to
the accidental coincidence between the collective energy and the
low-lying quasiparticle energy. As a result, the superfluidity of
the Fermi gas could be destroyed by the excitations of
quasiparticles. Nevertheless, the tendency of monotonic increase
of the predicted mode frequency in the BEC limit, as a natural
outcome of the mean-field theory, agrees qualitatively with the
experimental observations. This is to be contrasted with the
findings by Stringari \cite{stringari} which predict a decrease in
the transverse mode frequency towards the BEC limit.

In our calculations the collective mode frequencies in the trap
depend mainly on the dependence of the chemical potential on
density through an effective power-law exponent. We have not taken
account of non-mean-field corrections to the compressibility,
which were discussed in Ref. \cite{stringari}. These can be
estimated in a perturbative way for $\kappa\rightarrow \pm
\infty$, and can be shown to be very small near the BEC limit from
the Lee-Huang-Yang correction term \cite{LHY} evaluated with an
estimate of the molecule-molecule coupling constant by Petrov {\it
et al.} \cite{petrov}. The Hartree term, on the other hand, tends
to emphasize the dip shown in Fig. 2 near the BCS limit.

The calculated collective modes as functions of the anisotropy
parameter $\lambda $ are shown in Fig. 3 for four values of the
coupling strength. The sensitivity of the transverse mode to the
coupling is preserved for all values of $\lambda $, while the
axial mode is less sensitive to the interactions. For a spherical
trap with ($\lambda =1$) the breathing modes are known as the
monopole and the quadrupole mode and we have $\omega _{+}/\omega
_{\bot }=\sqrt{2+3\bar {\gamma}}$ and $\omega _{-}/\omega
_z=\sqrt{2} $. In the oblate limit ($\lambda \gg 1$) the
frequencies are $\omega_{+}/\omega _z=\sqrt{2+\bar {\gamma}}$ and
$\omega _{-}/\omega _{\bot }=\sqrt{ \left( 4+6\bar {\gamma}
\right) /\left( 2+\bar {\gamma} \right) }$.

\begin{figure}[tbp]
\centerline{\includegraphics[width=5.5cm,angle=-90,clip=]{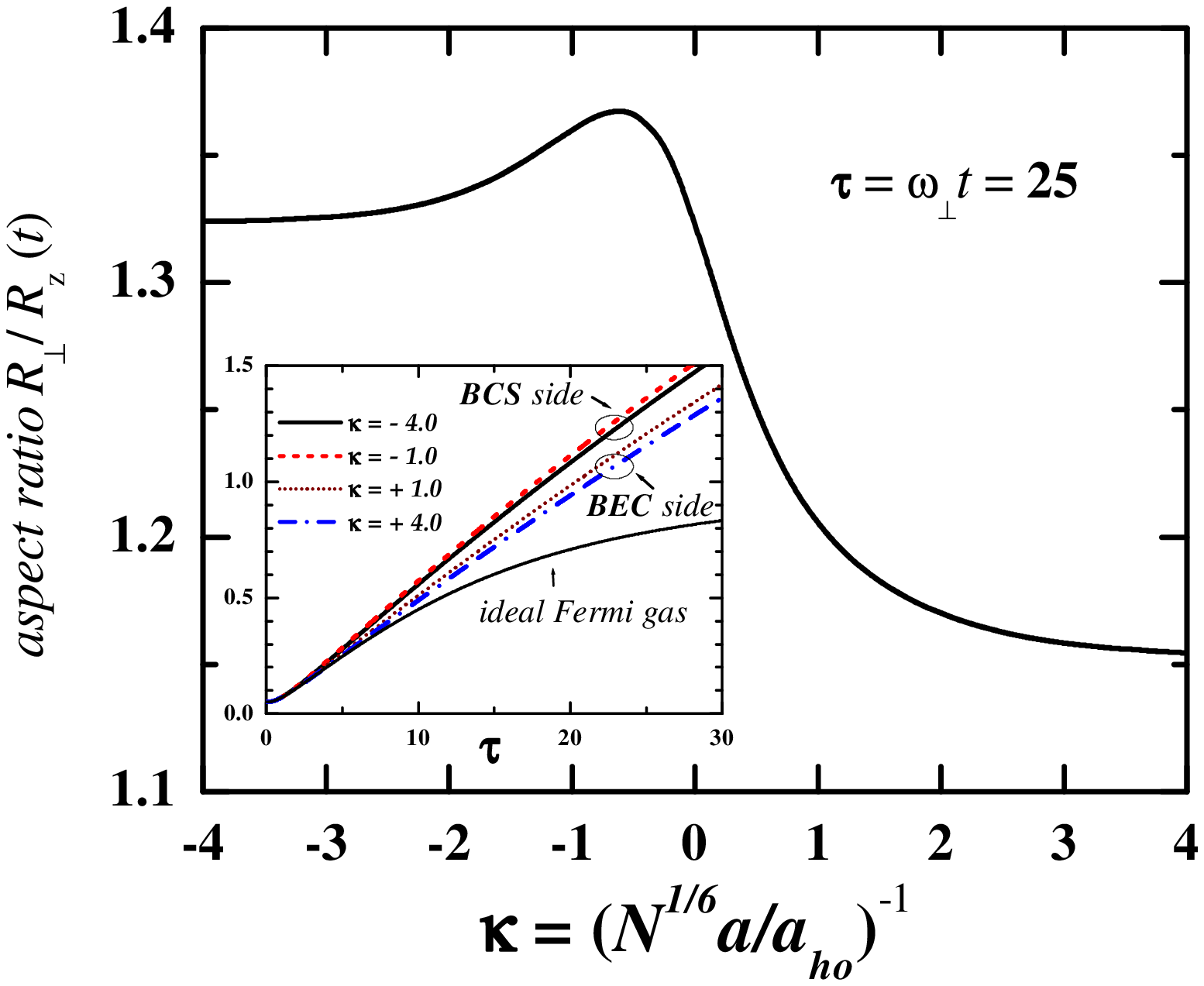}}
\caption{(color online) The aspect ratio $\lambda b_{\perp }/b_z$
as a function of $\kappa$ at a fixed time $\tau =\omega_{\bot } t
= 25$. The inset shows the aspect ratio as a function of the
dimensionless time $\tau$ for four values of the coupling
parameters $\kappa$ as indicated. For comparison the expansion of
an ideal gas is also shown in the inset by a thin solid line. The
other parameters are $N=2\times 10^5$ and $\lambda =0.05$.}
\label{fig4}
\end{figure}

\noindent {\it Expansion of the Fermi gas}. --- Finally we turn to study the
expansion of the Fermi gas after suddenly switching off the
confining potential, by setting $ \omega _\alpha (t>0)=0$. We
follow the time evolution of the aspect ratio $R_{\perp
}(t)/R_z(t)=\lambda b_{\perp }\left( t\right) /b_z\left( t\right)
$ by numerically solving Eq. (\ref{sfinal}) with the initial
configurations $b_\alpha (0)=1$ and $\dot{b}_\alpha (0)=0 $.

The results are shown in Fig. 4. The aspect ratio of the cloud at
a given expansion time shows non-monotonic behavior as a function
of the coupling strength $\kappa$, similarly to what is observed
in the collective modes frequencies. This reflects again the
emergence of important interaction effects in the hydrodynamic
behavior.

\noindent {\it Conclusions}. --- In summary, we have presented a
microscopic theory of the dynamics of a superfluid Fermi gas
through the whole BEC-BCS crossover. Our analysis is based on the
equation of state of the gas in a mean-field approach. For a gas
under anisotropic confinement we have found that both the
frequencies of the breathing modes and the aspect ratio of the
expanding cloud show non-monotonic behavior as functions of the
coupling constant. We have also made quantitative contact with
current experiments.

This work has been partially supported by INFM under the
PRA-Photonmatter program. X.-J. L. was supported by NSF-China
under Grant No. 10205022 and by the National Fundamental Research
Program under Grant No. 001CB309308. We thank Prof. J. Thomas for
useful discussions on the experimental data. A.M. thanks Prof. G.
V. Shlyapnikov for useful discussions and the LPTMS Orsay, where
this work was completed, for their hospitality.

\end{document}